%Paper: hep-ph/9405414
%From: kharzeev@axcrnc.cern.ch
%Date: Mon, 30 May 1994 18:26:25 +0200
%Date (revised): Mon, 6 Jun 1994 15:38:25 +0200

\def\J{$J/\psi$}
\def\j{J/\psi}
\def\P{$\psi'$}

\def\U{$\Upsilon$}

\def\c{c{\bar c}}
\def\b{b{\bar b}}

\def\F{$\Phi$}
\def\f{\Phi}

\def\t{\tau}

\def\Q{Q{\bar Q}}

\def\l{\lambda}
\def\e{\epsilon}

\def\lsim{\raise0.3ex\hbox{$<$\kern-0.75em\raise-1.1ex\hbox{$\sim$}}}
\def\gsim{\raise0.3ex\hbox{$>$\kern-0.75em\raise-1.1ex\hbox{$\sim$}}}

\newcount\REFERENCENUMBER\REFERENCENUMBER=0
\def\REF#1{\expandafter\ifx\csname RF#1\endcsname\relax
               \global\advance\REFERENCENUMBER by 1
               \expandafter\xdef\csname RF#1\endcsname
                   {\the\REFERENCENUMBER}\fi}
\def\reftag#1{\expandafter\ifx\csname RF#1\endcsname\relax
               \global\advance\REFERENCENUMBER by 1
               \expandafter\xdef\csname RF#1\endcsname
                      {\the\REFERENCENUMBER}\fi
             \csname RF#1\endcsname\relax}
\def\ref#1{\expandafter\ifx\csname RF#1\endcsname\relax
               \global\advance\REFERENCENUMBER by 1
               \expandafter\xdef\csname RF#1\endcsname
                      {\the\REFERENCENUMBER}\fi
             [\csname RF#1\endcsname]\relax}
\def\refto#1#2{\expandafter\ifx\csname RF#1\endcsname\relax
               \global\advance\REFERENCENUMBER by 1
               \expandafter\xdef\csname RF#1\endcsname
                      {\the\REFERENCENUMBER}\fi
           \expandafter\ifx\csname RF#2\endcsname\relax
               \global\advance\REFERENCENUMBER by 1
               \expandafter\xdef\csname RF#2\endcsname
                      {\the\REFERENCENUMBER}\fi
             [\csname RF#1\endcsname--\csname RF#2\endcsname]\relax}
\def\refs#1#2{\expandafter\ifx\csname RF#1\endcsname\relax
               \global\advance\REFERENCENUMBER by 1
               \expandafter\xdef\csname RF#1\endcsname
                      {\the\REFERENCENUMBER}\fi
           \expandafter\ifx\csname RF#2\endcsname\relax
               \global\advance\REFERENCENUMBER by 1
               \expandafter\xdef\csname RF#2\endcsname
                      {\the\REFERENCENUMBER}\fi
            [\csname RF#1\endcsname,\csname RF#2\endcsname]\relax}
\def\refss#1#2#3{\expandafter\ifx\csname RF#1\endcsname\relax
               \global\advance\REFERENCENUMBER by 1
               \expandafter\xdef\csname RF#1\endcsname
                      {\the\REFERENCENUMBER}\fi
           \expandafter\ifx\csname RF#2\endcsname\relax
               \global\advance\REFERENCENUMBER by 1
               \expandafter\xdef\csname RF#2\endcsname
                      {\the\REFERENCENUMBER}\fi
           \expandafter\ifx\csname RF#3\endcsname\relax
               \global\advance\REFERENCENUMBER by 1
               \expandafter\xdef\csname RF#3\endcsname
                      {\the\REFERENCENUMBER}\fi
[\csname RF#1\endcsname,\csname RF#2\endcsname,\csname
RF#3\endcsname]\relax}
\def\refand#1#2{\expandafter\ifx\csname RF#1\endcsname\relax
               \global\advance\REFERENCENUMBER by 1
               \expandafter\xdef\csname RF#1\endcsname
                      {\the\REFERENCENUMBER}\fi
           \expandafter\ifx\csname RF#2\endcsname\relax
               \global\advance\REFERENCENUMBER by 1
               \expandafter\xdef\csname RF#2\endcsname
                      {\the\REFERENCENUMBER}\fi
            [\csname RF#1\endcsname,\csname RF#2\endcsname]\relax}
\def\Ref#1{\expandafter\ifx\csname RF#1\endcsname\relax
               \global\advance\REFERENCENUMBER by 1
               \expandafter\xdef\csname RF#1\endcsname
                      {\the\REFERENCENUMBER}\fi
             [\csname RF#1\endcsname]\relax}
\def\Refto#1#2{\expandafter\ifx\csname RF#1\endcsname\relax
               \global\advance\REFERENCENUMBER by 1
               \expandafter\xdef\csname RF#1\endcsname
                      {\the\REFERENCENUMBER}\fi
           \expandafter\ifx\csname RF#2\endcsname\relax
               \global\advance\REFERENCENUMBER by 1
               \expandafter\xdef\csname RF#2\endcsname
                      {\the\REFERENCENUMBER}\fi
            [\csname RF#1\endcsname--\csname RF#2\endcsname]\relax}
\def\Refand#1#2{\expandafter\ifx\csname RF#1\endcsname\relax
               \global\advance\REFERENCENUMBER by 1
               \expandafter\xdef\csname RF#1\endcsname
                      {\the\REFERENCENUMBER}\fi
           \expandafter\ifx\csname RF#2\endcsname\relax
               \global\advance\REFERENCENUMBER by 1
               \expandafter\xdef\csname RF#2\endcsname
                      {\the\REFERENCENUMBER}\fi
        [\csname RF#1\endcsname,\csname RF#2\endcsname]\relax}
\def\refadd#1{\expandafter\ifx\csname RF#1\endcsname\relax
               \global\advance\REFERENCENUMBER by 1
               \expandafter\xdef\csname RF#1\endcsname
                      {\the\REFERENCENUMBER}\fi \relax}

%
%DEFINE JOURNAL NAMES

\def\NP{{ Nucl.\ Phys.\ }}
\def\PL{{ Phys.\ Lett.\ }}
\def\PR{{ Phys.\ Rev.\ }}

\def\PRL{{ Phys.\ Rev.\ Lett.\ }}

\def\ZP{{ Z.\ Phys.\ }}

%

% "New" FINAL FILE - 2.6.1994
\magnification=1200
\hsize=16.0truecm
\vsize=24.5truecm
\baselineskip=13pt
\pageno=0
{}~~~
\hfill CERN-TH.7274/94\par
\hfill BI-TP 94/24\par
\vskip 2.5truecm
\centerline{\bf QUARKONIUM INTERACTIONS IN HADRONIC MATTER}
\vskip 1.5 truecm
\centerline{D.\ Kharzeev and H.\ Satz}
\bigskip\medskip
\centerline{Theory Division, CERN, CH-1211 Geneva, Switzerland}
\centerline{and}
\centerline{Fakult\"at f\"ur Physik, Universit\"at Bielefeld,
D-33501 Bielefeld, Germany}
\vskip 2 truecm
\centerline{\bf Abstract:}
\medskip
The cross section for the \J~and \U~interaction with light hadrons
is calculated in short-distance QCD, based on the large heavy quark
mass and the resulting large energy gap to open charm or beauty. The low
energy form of the cross section is determined by the gluon structure
functions at large $ x$; hence it remains very small until quite high
energies. This behaviour is experimentally confirmed by charm
photoproduction data. It is shown to exclude \J~absorption in confined
hadronic matter of the size or density attainable in nuclear
collisions; in contrast, the harder gluon spectrum in deconfined
matter allows break-up interactions.
\par\vfill\noindent
CERN-TH.7274/94\par\noindent
BI-TP 94/24\par\noindent
May 1994
\eject
\refadd{Peskin}
\refadd{Bhanot}
\refadd{SVZ}
\refadd{Novikov}
\refadd{Matsui}
\refadd{Holmes}
\refadd{Nash}
\refadd{Rossi}
\refadd{Ali}
The interaction of heavy quark-antiquark resonances (quarkonia) with
the ordinary light hadrons plays an important role both in the
dynamics and in the thermodyna-\break mics of strong interaction
physics.
For QCD dynamics, it is important since the small quarkonium size probes
the short-distance aspects of the big light hadrons and thus makes
a parton-based calculation of the overall cross section possible
\refto{Peskin}{Novikov}. In QCD thermodynamics,
quarkonia can be used as a probe for deconfinement \ref{Matsui},
provided their interaction in dense confined matter
can be distinguished from that in a quark-gluon plasma.
\par
In this letter we will first recall the QCD analysis of quarkonium
interactions with light hadrons. It concludes that the
small size of quarkonia, combined with the rather large mass gap to
open charm or beauty, strongly inhibits their break-up by low energy
collisions with light hadrons \ref{Bhanot}; the total quarkonium-hadron
cross section attains a constant asymptotic value
only at very high energies, compared to corresponding cross sections for
the interaction between light hadrons. This slowly rising form of the
cross section is derived from an operator product
expansion with ensuing sum rules and becomes quite transparent in parton
language. It is experimentally well supported by applying the
corresponding analysis to data for charm photoproduction
\refto{Holmes}{Ali}.
\par
We then use this result to show that a slow \J~or \U~cannot be
absorbed (i.e., broken up into a $D \bar{D}$) in hadronic matter of
the volume and density which can be obtained in nuclear collisions.
The essential point is
here that absorption requires interactions with sufficiently
hard gluons, and the gluon distribution within hadrons does not
provide these for matter at meaningful temperatures. Hence
absorption in confined hadronic matter is excluded as a possible cause
of the \J~suppression observed in nucleus-nucleus collisions. On the
other hand, we find that quarkonia can be broken in
deconfined matter, which contains much harder gluons. This provides
a dynamical basis of quarkonium suppression by colour screening
\refs{Matsui}{Mehr}.
\par
The QCD analysis of quarkonium interactions applies
to heavy and strongly bound quark-antiquark
states \ref{Bhanot}; therefore we here restrict ourselves to the
lowest $\c$ and $\b$ vector states \J~and \U, which we denote
generically by \F, following the notation of \ref{Bhanot}.
For such states, both the masses $m_Q$ of the constituent
quarks and the binding energies $\e_0(\f) \simeq (2M_{(Qq)}
- M_{\f}$) are much larger
than the typical scale $\Lambda_{\rm QCD}$ for non-perturbative
interactions; here $(Qq)$ denotes the lowest open charm or beauty
state. In $\f-h$ interactions, as well as in \F-photoproduction,
$\gamma h \to \f h$, we thus only probe a
small spatial region of the light hadron $h$; these processes are much
like deep-inelastic lepton-hadron scattering, with large $m_Q$ and
$\e_0$ in place of the large virtual photon mass $\sqrt{-q^2}$.
As a result, the calculation of \F-photoproduction and of
absorptive $\f-h$ interactions
can be carried out in the short-distance formalism
of QCD. Just like deep-inelastic leptoproduction, these reactions
probe the parton structure
of the light hadron, and so the corresponding cross sections can
be calculated in terms of parton interactions and structure
functions.
\par
In the following, we shall first sketch the theoretical basis which
allows quarkonium interactions with light hadrons to be treated
by the same techniques as used in deep-inelastic lepton-hadron
scattering or in the photoproduction of charm. We show the derivation
for the sum rules which relate the absorptive $\f-h$ cross section
to hadronic gluon structure functions \refs{Peskin}{Bhanot}.
This relation given, we calculate explicitly the energy dependence of
the cross section. Readers only interested in this
behaviour can therefore go immediately to eq.\ (24).
\par
Consider the amplitude for forward scattering of a virtual
photon on a nucleon,
$$
F(s, q^2) \sim
i\int d^4x e^{iqx} <N|T\{J_{\mu}(x)J_{\nu}(0)\}|N>. \eqno(1)
$$
In the now standard application of QCD to
deep-inelastic scattering one exploits the fact that at large
spacelike photon momenta $q$ the amplitude is dominated by small
distances of order $1/\sqrt{-q^2}$ (Fig. 1a).
The Wilson operator product expansion then
allows the evaluation of the amplitude at the unphysical point
$pq\to 0$, where $p$ is the four-momentum of the nucleon.
Since the imaginary part of the amplitude (1) is proportional
to the experimentally observed structure functions of deep-inelastic
scattering, the use of dispersion relations relates the value
of the amplitude at $pq\to 0$ point to the integrals over the
structure functions, leading to a set of dispersion sum rules
\ref{Gross}.
The parton model can be considered then as a particularly useful
approach satisfying these sum rules.
\par
In the case $J_{\mu} = \bar{Q} \gamma_{\mu} Q$, i.e., when
vector electromagnetic current in eq.\ (1) is that
of a heavy quark-antiquark pair, large momenta $q$ are not needed
to justify the use of perturbative methods.
Even if $q\sim 0$, the small space-time scale of $x$ is set by the
mass of the charmed quark, and the characteristic distances
which are important in the correlator (1) are of the order of
$1 /2m_Q$ (Fig.\ 1b). In \refs{SVZ}{Novikov}, this observation was used
to
derive sum rules for charm photoproduction in a manner quite similar to
that used for deep-inelastic scattering.
\par
In the interaction of quarkonium with light hadrons, again the
small space scale is set by the
mass of the heavy quark, and the characteristic distances involved
are of the order of quarkonium size, i.e., smaller than the
non-perturbative hadronic scale $\Lambda_{QCD}^{-1}$.
Moreover, since heavy quarkonium and light hadrons do not have
quarks in common, the only allowed exchanges are purely gluonic.
However, the smallness of spatial size is not enough to justify the use
of perturbative expansion \ref{Bhanot}. Unlike in the case of
\F-photoproduction, heavy quark lines now appear in the initial
and final states (see Fig.\ 1c), so that the $(\Q)$ state can emit and
absorb gluons at points along its world line widely separated in
time. These gluons must be hard enough to interact
with a compact colour singlet state (colour screening leads to a
decoupling of soft gluons with the wavelengths larger than the
size of the \F); however, the interactions among the gluons
can be soft and nonperturbative. We thus have to assure that the
process is compact also in time.
Since the absorption or emission of a gluon turns
a colour singlet quarkonium state into a colour octet,
the scale which regularizes the time correlation of such processes
is by the quantum-mechanical uncertainty
principle just the mass difference between the colour-octet and colour-
singlet states of quarkonium: $\tau_{c} \sim 1/ (\e_8 - \e_1)$.
The perturbative Coulomb-like piece of the heavy quark-antiquark
interaction
$$
V_k(r)=-g^2 {c_k \over {4\pi r}} \eqno(2)
$$
is attractive in the colour singlet ($k=1$) and repulsive in the
colour-octet ($k=8$) state; in SU(N) gauge theory
$$
V_1=-{g^2 \over {8\pi r}} {N^2-1 \over N}, \eqno(3a)
$$
$$
V_8={g^2 \over {8\pi r}} {1 \over N}. \eqno(3b)
$$
To leading order in $1/N$, the mass gap between the singlet and octet
states is therefore just the binding energy of the heavy
quarkonium $\e_0$, and the characteristic correlation time for
gluon absorption and emission is
$$
\tau_c \sim 1/\e_0. \eqno(4)
$$
Although the charm quark is not heavy enough to ensure a pure Coulomb
regime even for the lowest $c\bar{c}$ bound states ($\eta_c$ and $\j$),
the mass gap determined from the observed value of open charm threshold
clearly shows that $\tau_c < \Lambda_{QCD}$. For the \U, the interaction
is in fact essentially Coulomb-like and the mass gap to open beauty
is even larger than for charm. One therefore expects to be able to treat
quarkonium interactions with light hadrons by the same QCD methods
that are used in deep-inelastic scattering and charm photoproduction.
\par
We thus use the operator product
expansion to compute the amplitude of heavy quarkonium
interaction with light hadrons,
$$
F_{\f h} = i\int d^4x e^{iqx} \langle h|T\{J(x)J(0)\}|h \rangle =
\sum_n c_n(Q,m_Q) \langle O_n \rangle ,
\eqno(5)
$$
where the set $\{O_n\}$ includes all local gauge invariant operators
expressible in terms of gluon fields; the matrix elements $\langle
O_n \rangle$ are taken between the initial and final light-hadron
states. The coefficients $c_n$ are computable perturbatively
\ref{Peskin} and process-independent. As noted above, in
deep-inelastic scattering
the expansion (5) is useful only in the vicinity of the point
$pq\to 0$. The same is true for the case of quarkonium interaction
with light hadrons. As shown in \ref{Bhanot}, the expansion (5) can
therefore be rewritten as an expansion in the variable
$$
\lambda = {pq \over M_{\f}} = {(s-M_{\f}^2 - M_h^2)\over 2M_{\f}}
\simeq {(s-M_{\f}^2)\over 2M_{\f}}
\eqno(6)
$$
where $M_h$ is the mass of the light hadron;
the approximate equality becomes valid the heavy quark limit.
For the lowest 1S quarkonium state one then obtains
$$
F_{\f h} = r_0^3\ \e_0^2\ \sum_{n=2}^{\infty}
d_n \langle O_n \rangle
\left({\lambda \over \e_0}\right)^n, \eqno(7)
$$
where $r_0$ and $\e_0$ are Bohr radius and binding energy of the
quarkonium, and the sum runs over even values of $n$ to ensure the
crossing symmetry of the amplitude.
The most important coefficients $d_n$ were computed
in \ref{Peskin} to leading order in $g^2$ and $1/N$.
\par
Since the total $\f-h$ cross section $\sigma_{\f h}$ is proportional
to the imaginary part of the amplitude $F_{\f h}$, the dispersion
integral over $\l$ leads to the sum rules
$$
{2\over \pi}\int^{\infty}_{\l_0} d\l~\l^{-n} \sigma_{\f h}(\l)
= r_0^3~ \e_0^2~ d_n \langle O_n \rangle \left( {1\over \e_0}
\right)^n.
\eqno(8)
$$
Eq.\ (7) provides only the inelastic intermediate states in the
unitarity
relation, since direct elastic scattering leads to contributions of
order $r_0^6$. Hence the total cross section in eq.\ (8) is due to
absorptive interactions only \ref{Bhanot}, and the integration
in eq.\ (8) starts at a lower limit $\l_0 > M_h$.
 Recalling now the expressions for radius and binding energy of
1S Coulomb bound states of a heavy quark-antiquark pair,
$$
r_0 = \left( {16\pi \over {3 g^2}} \right) {1 \over m_Q}, \eqno(9)
$$
$$
\e_0 = \left( {3 g^2 \over {16 \pi}} \right)^2 m_Q, \eqno(10)
$$
and using the coefficients $d_n$ from
\ref{Peskin}, it is possible \ref{Bhanot} to rewrite these sum rules
in the form
$$
\int_{\l_0}^{\infty} { d\lambda \over \l_0} \left( {\lambda \over
\l_0} \right)^{-n}
\sigma_{\f h} (\lambda) = 2 \pi^{3/2}\ \left(16 \over 3 \right)^2\
{{\Gamma \left(n + {5\over 2} \right)} \over {\Gamma (n + 5)}}\
\left( {16\pi \over {3g^2}}\right)\
{1 \over m_Q^2}\ \langle O_n \rangle, \eqno(11)
$$
with $\l_0/\e_0 \simeq 1$ in the heavy quark limit.
The contents of these sum rules become more transparent
in terms of the parton model. In parton language,
the expectation values $\langle O_n \rangle$ of the operators composed
of gluon fields can be expressed as Mellin
transforms \ref{Parisi} of the gluon structure function of the light
hadron, evaluated at the scale $Q^2=\e_0^2$,
$$
\langle O_n \rangle = \int_0^1 dx\ x^{n-2} g(x, Q^2 = \e_0^2).
\eqno(12)
$$
Defining now
$$
y = {\l_0 \over \lambda} , \eqno(13)
$$
we can reformulate eq.\ (11) to obtain
$$
\int_0^1 dy\ y^{n-2} \sigma_{\f h}(\l_0 / y) = I(n)\
\int_0^1 dx\ x^{n-2} g(x, Q^2 = \e_0^2), \eqno(14)
$$
with $I(n)$ given by
$$
I(n) =  2 \pi^{3/2}\ \left(16 \over 3 \right)^2\
{{\Gamma \left(n + {5\over 2} \right)} \over {\Gamma (n + 5)}}\
\left( {16\pi \over {3g^2}}\right)\ {1 \over m_Q^2}. \eqno(15)
$$
Eq.\ (14) relates the $\f-h$ cross section to the gluon structure
function. To get a first idea of this relation, we neglect the
$n$-dependence of $I(n)$ compared to that of $\langle O_n \rangle$; then
we conclude that
$$
\sigma_{\f h}(\l_0 / x) \sim g(x, Q^2=\e_0^2), \eqno(16)
$$
since all order Mellin transforms of these quantities are equal up to a
constant. From eq.\ (16) it is clear that
the energy dependence of the $\f-h$ cross section is entirely
determined by the $x-$dependence of the gluon structure function.
The small $x$ behaviour of the structure function governs
the high energy form of the cross section, and the hard tail
of the gluon structure function for $x \to 1$ determines
the energy dependence of $\sigma_{\f h}$ close to the
threshold.
\par
To obtain relation (16), we have neglected the $n$-dependence of the
function $I(n)$. Let us now try to find a more accurate solution of the
sum rules (13). We are primarily interested in the energy region not
very far from the inelastic threshold, i.e.,
$$
(M_h + \e_0)\ \lsim \l\ \lsim\ 5\ {\rm GeV}, \eqno (17)
$$
since we want to calculate in particular the absorption of \F's
in confined hadronic matter. In such an environment,
the constituents will be hadrons with momenta of at most
a GeV or two. A usual hadron ($\pi,~\rho$, nucleon) of 5 GeV momentum,
incident on a \J~at rest, leads to $\sqrt s \simeq 6$ GeV, and this
corresponds to $\l \simeq 5$ GeV.
\par
{}From what we learned above, the energy region corresponding to the range
(17) will be determined by the gluon structure function at
values of $x$ not far from unity. There
the $x$-dependence of $g(x)$ can be well described by a power law
$$
g(x) = g_2\ (k+1)\ (1-x)^k, \eqno(18)
$$
where the function (18) is normalized so that the second moment (12)
gives the fraction $g_2$ of the light hadron momentum carried
by gluons, $ <O_2> = g_2 \simeq 0.5$.
This suggests a solution of the type
$$
\sigma_{\f h}(y) = a (1-y)^{\alpha}, \eqno(19)
$$
where $a$ and $\alpha$ are constants to be determined.
Substituting (18) and (19) into the sum rule (13) and performing the
integrations, we find
$$
a\ {{\Gamma(\alpha+1)} \over {\Gamma(n+\alpha)}} =
\left({2 \pi^{3/2} g_2 \over m_Q^2}\right)\left(16 \over 3 \right)^2
\left( {16\pi \over {3g^2}}\right)
{\Gamma(n+{5 \over 2}) \over \Gamma(n+5)} {\Gamma(k+2) \over
\Gamma(k+n)}.
\eqno(20)
$$
We are interested in the region of low to moderate
energies; this corresponds to relatively large $x$, to which higher
moments are particularly sensitive.
Hence for the
range of $n$ for which eq.\ (5) is valid \ref{Novikov},
$n~\lsim~8$, the essential
$n$-dependence is contained in the $\Gamma$-functions.
For $n~\gsim~4$, eq.\ (20) can
solved in closed form by using an appropriate approximation for the
$\Gamma$-functions. We thus obtain
$$
a\ {\Gamma(\alpha +1) \over \Gamma(k+2)} \simeq {\rm const.}\
n^{\alpha - k -5/2}. \eqno(21)
$$
Hence to satisfy the sum rules (14), we need
$$
\alpha=k+{5 \over 2}~~~~~~~
a = {\rm const.}\ {\Gamma(k+2) \over \Gamma(k+{7 \over 2})}. \eqno(22)
$$
Therefore the solution of the sum rules (13) for
moderate energies $\lambda$ takes the form\footnote{*}{The same
functional form for the energy dependence was also obtained in
\ref{Bhanot} through parton model arguments.}
$$
\sigma_{\f h}(\lambda) = 2 \pi^{3/2} g_2  \left(16 \over 3 \right)^2
\left( {16\pi \over {3g^2}}\right){1 \over m_Q^2}
{\Gamma(k+2) \over \Gamma(k+{7 \over 2})}
\left(1-{\l_0 \over \l}\right)^{k+5/2}. \eqno(23)
$$
To be specific, we now consider the \J-nucleon interaction.
Setting $k=4$ in accord with quark counting rules, using
$g_2\simeq 0.5$ and expressing the strong coupling $g^2$ in terms of
the binding energy $\e_0$ (eq.\ 10),
we then get from eq.\ (23) the energy dependence of
the $\j~\!N$ total cross section
$$
\sigma_{\j N}(\l) \simeq 2.5\ {\rm mb} \times
\left(1-{\l_0 \over \l}\right)^{6.5}, \eqno(24)
$$
with $\l$ given by eq.\ (6) and $\l_0 \simeq (M_N+\e_0)$.
This cross section rises very slowly from threshold, as shown in Fig.\
2; for $P_N \simeq 5 $ GeV,
it is around 0.1 mb, i.e., more than an
order of magnitude below its asymptotic value. This clearly shows that
hadronic matter at meaningful densities and for the volumes
relevant for nuclear collisions is not efficient in breaking up
quarkonium states. To illustrate this, we note that in a volume
of the size of a uranium nucleus, the average path length for a \J~is
$L\simeq (3/4)(1.2~A^{1/3})\simeq 5.6$ fm. For matter of standard
nuclear density $n_0=0.17$ fm$^{-3}$, a cross section
$\sigma_{\j N}=0.1$ mb leads to a survival
probability $S={\rm exp}(-\sigma_{\j N}~n_0~L)\simeq 0.99$, and
even at five times standard nuclear density, we still have a survival
probability of more than 95\%. --
We should note that the high energy cross section of 2.5 mb in eq.\ (24)
is {\sl calculated} in the short-distance formalism of QCD and
determined numerically by the values of $m_c$
and $\e_0$.\footnote{*}{An
alternative and in principle equivalent approach
is to calculate the $\f-N$ cross section by perturbative QCD
\ref{Nason}. The sum rule formalism used here has the advantage
of providing the total cross section directly in terms of the largely
model-independent scales $m_Q,~\e_0$ and the fraction $g_2$.}
It
agrees very well with the 2 - 3 mb obtained from photoproduction
data via vector meson dominance \ref{Holmes}.
\par
For sufficiently heavy quarkonia, the calculation of the $\f-h$
cross section
as presented is firmly based on short-distance QCD. To check its
applicability for charmonium, experimental tests are certainly
important. This is indeed possible, since a completely analogous
derivation \ref{Novikov} provides the cross section of charm
photoproduction, $\gamma~\!N\to \c~X$. Instead of eq.\ (24), we
then get
$$
\sigma_{\gamma N \to \c}(\nu) \simeq 1.2 ~\mu{\rm b} \times
\left(1-{\nu_0 \over \nu}\right)^{4}, \eqno(25)
$$
in terms of the scaling variable $\nu=(s-M_N^2)/2$, with
$\nu_0=[(M_\j+M_N)^2 - M_N^2]/2$. The exponent in eq.\ (25) is again
determined by the gluon structure function; it differs from that in
eq.\ (24) because of the absence of the bound state
Coulomb wave function in computing the coefficient function for
$\c$ photoproduction.
In Fig.\ 3 we compare the result (25) to a summary of data
\refto{Holmes}{Ali}. Since eq.\ (25) contains the contributions of
both open and hidden charm, we have reduced it in Fig.\ 3
by 25\% to account for the absence of elastic
shadow scattering in the data. It is seen to agree quite well
with the experimental behaviour, which in particular shows clearly the
small cross section in the threshold region.
\par
A similar test is in principle provided by \J-photoproduction,
$\gamma~\! N\to \j~\!N$. The forward scattering amplitude of this
process will give the total charm photoproduction cross section
$\sigma_{\gamma N\to {\c}}$, provided vector meson dominance holds and
the real part of the amplitude can be neglected.
Near threshold presumably neither is true, and so (Fig.\ 4)
the result of our calculation falls there below the data \ref{Holmes},
even though also here the cross section attains its high energy
value very slowly.
\par
The calculation of the total $\f h$ cross section as given above
includes all possible quarkonium break-up mechanisms and
becomes exact in the heavy quark limit. It is interesting, however,
to consider also non-perturbative mechanisms made possible by
the finiteness of the charmed quark mass.
Here the ``subthreshold" process \ref{Khar}
$$
J/\Psi + N \rightarrow \Lambda_c + \bar{D} \eqno(26)
$$
deserves particular attention, since it can occur even at very small
energies. Apart from the perturbative contribution calculated above,
the amplitude for reaction (26)
can acquire a non-perturbative piece, corresponding to
the tunneling of the $c$-quark from the charmonium potential well
over a large distance $ R_t \sim \Lambda_{QCD}^{-1}$ into the
potential well of the nucleon. One can roughly estimate the probability
of this process.
The time $\t_t$ required for the tunneling over the distance $R_t$
is $\t_t \sim \Lambda_{QCD}^{-1}$. On the other hand, the energy
fluctuation needed to overcome the potential barrier of the height
$\epsilon_0$ has according to the uncertainty relation
a much shorter lifetime,
of order $\t_f \sim \epsilon_0^{-1}$.
The probability for tunneling is thus expected to be
$$
W_t \simeq {\rm exp}(-\t_t / \t_f) \simeq
{\rm exp} (-\e_0 / \Lambda_{QCD}). \eqno(27)
$$
This probability vanishes as expected in the heavy quark limit, since
$\epsilon_0 \sim m_Q$ (see eq.\ (10)). For the \J,
the probability (27) is already quite small (about $0.02$),
and hence the corresponding non-perturbative contribution
is unlikely to affect
the total quarkonium absorption cross section in any significant way.
\par
\refadd{Alde}
\refadd{Khar2}
\refadd{Khar1}
\refadd{Ftac}
\refadd{Blaizot}
\refadd{Vogt}
Since we have concluded that quarkonia cannot be broken up in hadronic
matter, we should
comment briefly on the implication of our results for $h A$ collisions.
In presently available data, the produced quarkonia (\J,~\P,~\U)
have very high momenta in the rest frame of the target nucleus.
Hence they are not yet physical resonances when they leave the nuclear
medium; as an experimental consequence of this, \J~and \P~, although
very different in size, suffer the same nuclear suppression \ref{Alde}.
The observed distributions for these nascent quarkonia are in fact
well described by a combination of quantum-mechanical
coherence effects (``shadowing") \ref{Khar2} and
colour octet interactions \ref{Khar1} of the $\Q$~system in the nucleus.
Thus there is so far no experimental information about the interaction
of fully formed physical quarkonia in the confined medium of a heavy
nucleus. To obtain such information, quarkonium production has to be
measured at large negative $x_F$ values \ref{Khar1}. For a hadron
beam incident on a nuclear target, this is difficult, since the
slow decay dileptons are hard to measure. The relevant experiment
is possible, however, for a nuclear beam incident on a hydrogen
target, so that the study of quarkonia in nuclear matter becomes
feasible with the advent of the Pb-beam at the CERN-SPS. Our analysis
predicts no (i.e., less than 5\%) suppression for the production of the
basic (1S) \J~in the range $-1 \leq  x_F  \lsim  -0.4$, in
contrast to some 25\% suppression obtained by using the
asymptotic $\f h$ cross section even in the threshold
region \refto{Ftac}{Vogt}. Since the mass of the \P~is
just at the open charm threshold, it is much easier to dissociate and
hence will suffer much stronger absorption. The same holds to a lesser
extent also for the $\chi$ states.
\par
In closing, we want to compare quarkonium interactions in confined and
in deconfined matter. In confined matter, the crucial gluon densities,
$g(x)\sim (1-x)^k$, lead to $\langle x \rangle = 1/(k+2)$,
which implies
$$
\langle p_g \rangle = {1\over k+2}~\langle P_h \rangle \eqno(28)
$$
for the average momentum of the gluons in terms of that of the hadron
incident on the quarkonium. For hadronic matter of temperature $T\simeq
200$ MeV, we thus get
$$
\langle p_g \rangle \simeq {3\over 5} T \simeq 0.12~{\rm GeV} \eqno(29)
$$
for the case of (massless) pions, with $P_{\pi}\simeq 3T$ and
$k=3$, and
$$
\langle p_g \rangle \simeq {1\over 3} \sqrt{2mT/\pi}
\simeq 0.12~{\rm GeV} \eqno(30)
$$
for nucleons, with $P_N \simeq 2\sqrt{2mT/\pi}$ and $k=4$.
It is clear that for gluons of such momenta it is difficult to break up
quarkonium, i.e., to overcome the threshold to open charm. In previous
absorption studies of charmonium suppression \ref{Gavin}, it was
suggested that this threshold is overcome in the collision of resonances
($\rho$, ...) with the \J, by making use of the difference between
resonance and pion mass, $M_{\rho} - 2 M_{\pi}$. Such arguments would be
correct for the interaction between light and large hadrons; but in
the interaction of the tiny
quarkonia with light hadrons, the short-distance
character due to the heavy quark mass and the large gap to open
charm or beauty makes the gluon distribution within the light hadron
the important feature, not its mass.
\par
In deconfined matter, on the other hand, we expect gluons in a medium
of temperature $T \simeq 0.2$ GeV to have an average momentum
$$
\langle p_g \rangle \simeq 3T \simeq 0.6~{\rm GeV}. \eqno(31)
$$
The impact of gluons with such a momentum spectrum can
easily overcome the 0.7 GeV
threshold to open charm and hence break up a \J~into $D$'s.
The essential feature of deconfinement in this dynamical view of
\J~suppression in a quark-gluon plasma \ref{Matsui} is thus the
hardening of the gluon spectrum.
\par \vfill
\centerline{\bf Acknowledgement:} \par
We thank J. Dias de Deus for helpful discussions.
The financial support of the German Research Ministry (BMFT), Contract 06
BI 721, is gratefully acknowledged.
\eject
\centerline{\bf References:}
\medskip
\item{\reftag{Peskin})}{M. E. Peskin, \NP B156 (1979) 365.}
\item{\reftag{Bhanot})}{G. Bhanot and M. E. Peskin, \NP B156 (1979)
391.}
\item{\reftag{SVZ})}{M. A. Shifman, A. I. Vainshtein and V. I. Zakharov,
 \PL 65B (1976) 255.}
\item{\reftag{Novikov})}
{ V. A. Novikov, M. A. Shifman, A. I. Vainshtein and V. I. Zakharov,
 \NP B136 (1978) 125.}
\item{\reftag{Matsui})}{T. Matsui and H. Satz, \PL B178 (1986) 416.}
\item{\reftag{Holmes})}
{S. D. Holmes, W. Lee and J. E. Wiss, Ann. Rev. Nucl. Part. Sci. 35
(1985) 397.}
\item{\reftag{Nash})}{T. Nash, {\sl Proc. of the Internat. Lepton/Photon
Symposium}, Cornell 1983, p. 329}
\item{\reftag{Rossi})}{L. Rossi, ``Heavy Quark Production", Genova
Preprint INFN/AE-91/16}
\item{\reftag{Ali})}{A. Ali, ``Heavy Quark Physics in Photo- and
Leptoproduction Processes at HERA and Lower Energies", DESY Preprint
93-105, 1933.}
\item{\reftag{Mehr})}{F. Karsch, M. T. Mehr and H. Satz, \ZP C37 (1988)
617.}
\item{\reftag{Gross})}{D. J. Gross and F. Wilczek, \PR D8 (1973) 3633
and \PR D9 (1974) 980;\hfill\break
H. Georgi and H. D. Politzer, \PR D9 (1974) 416.}
\item{\reftag{Parisi})}{G. Parisi, \PL 43B (1973) 207; 50B (1974) 367.}
\item{\reftag{Nason})}{See e.g. P. Nason et al., ``Heavy Flavour
Production in Perturbative QCD", CERN Preprint TH.7134/94, for
a recent survey.}
\item{\reftag{Khar})}{D. Kharzeev, \NP A558 (1993) 331c.}
\item{\reftag{Alde})}{D. M. Alde et al., \PRL 66 (1991) 133.}
\item{\reftag{Khar2})}{D. Kharzeev and H. Satz, \PL B, in
press.}
\item{\reftag{Khar1})}{D. Kharzeev and H. Satz, \ZP C60 (1993) 389.}
\item{\reftag{Ftac})}{J. Ft{\'a}{\v c}nik, P. Lichard and J.
Pi{\v s}\'ut, \PL B207 (1988) 194.}
\item{\reftag{Blaizot})}{J.-P. Blaizot and J.-Y. Ollitrault, \PL B217
(1989) 392.}
\item{\reftag{Vogt})}{S. Gavin, H. Satz, R. L. Thews and R. Vogt,  \ZP
C61 (1994) 351.}
\item{\reftag{Gavin})}{S. Gavin, M. Gyulassy and A. Jackson, \PL B207
(1988) 257.}
\vfill
\centerline{\bf Figure Captions:}
\leftskip 0.6 truecm
\medskip
\item{Fig.\ 1:}{Deep inelastic scattering (a), heavy flavour
photoproduction (b) and quarko-\break nium-nucleon interaction (c).}
\medskip
\item{Fig.\ 2:}{The energy dependence of the quarkonium-nucleon cross
section, normalised to its asymptotic value; also shown is the
dependence on the momentum $P_N$ of a nucleon incident on a \J~at rest.}
\medskip
\item{Fig.\ 3:}{The cross section of open charm photoproduction
\refto{Holmes}{Ali},
compared to the prediction of the short distance QCD analysis.}
\medskip
\item{Fig.\ 4:}{The total cross section for charm photoproduction
\ref{Holmes}, compared to the prediction of the short distance QCD
analysis, with the real part of the amplitude neglected.}
\eject
\bye